# On Weyl Nodes in Ferromagnetic Weyl Semimetal

*Udai Prakash Tyagi and Partha Goswami*[*]

*Deshbandhu College, University of Delhi, Kalkaji, New Delhi-110019,India*

[*]*E-mail of the corresponding author: physicsgoswami@gmail.com*

**ABSTRACT**

The ferromagnetic Weyl semimetals, such as $Co_3Sn_2S_2$, feature pairs of Weyl points characterized by the opposite chiralities. We model this type of semimetals by the inversion symmetry protected and the time reversal symmetry broken Bloch Hamiltonian. It involves terms representing the tunnelling effect, exchange field corresponding to the ferromagnetic order, chirality index of Weyl points with related energy parameters, and the angle formed by the spin magnetic moments and the axis perpendicular to the system-plane. While for the in-plane spin moment order the Weyl nodes are absent at some points of the first Brillouin zone, the bands of opposite chirality non-linearly cross each other with band inversion at Weyl points for the spin moment order along the perpendicular axis. The absence of linearity implies that the system is unable to host massless Weyl fermions. We also show that, in the absence of the exchange field, the incidence of the circularly polarized radiation leads to the emergence of a novel state with broken time reversal symmetry.

## 1. Introduction

A ferromagnetic Weyl semimetal (WSM) candidate, such as $Co_3Sn_2S_2$, features pairs of Weyl points **[1-18]** characterized by the opposite chiralities together with band inversion unlike a type-II WSM $MoTe_2$ **[19]**. The shandite compound $Co_3Sn_2S_2$, in the absence of a magnetic field, is known to be in ferromagnetic phase for temperature $T < T_c = 177$ K. However, at higher temperatures ($T > T_c$) it is a paramagnet. In their seminal theoretical investigation, Ozawa and Nomura **[17, 18]** had reported about this Weyl node feature starting with an archetypal model of $Co_3Sn_2S_2$. However, clarity regarding the nodes possessing opposite chirality was seemingly missing. They assumed that the angle formed by the spin moments and the axis perpendicular to the plane of the system was zero. Moreover, it was shown that this ferromagnetic compound, with magnetic kagome lattice, hosts massless Weyl fermions and gives rise to strong intrinsic quantum anomalous Hall (QAH) effect.

In this study, in order to examine the Weyl nodes of opposite chirality feature, we have started with a much simpler model of ferromagnetic WSM (FMWSM), which is a variant of that in ref. **[20,21]**. The essential ingredients of the model are terms representing the tunnelling effect, the exchange field ($\Delta$) representing the ferromagnetic order, the angle formed by the spin moments and the axis perpendicular to the plane, and the chirality index ($\xi = \pm 1$) of the expected Weyl nodes and the related energy parameter ($\zeta$) determining the shift of the nodes. It should be emphasized that the parameters ($\xi, \zeta$), which were not taken into consideration in ref.[17,18], are the crucial parameters. It is due to these parameters that we have been able to show the presence of the Weyl points of *opposite chirality* in the band structure of FMWSM with band inversion in a certain parameter range, when the angle formed by the spin moments and the axis perpendicular to the plane of the system is zero. We found, however, the lack the presence of the Weyl nodes at some points in the Brillouin zone (BZ)in the case of the in-plane spin order. In Sect. 2, we have calculated the anomalous Hall conductivity(AHC) based on this model. The surface exposition to circularly polarized radiation (CPR) in the case of WSM thin films has been rarely explored so far. We also report the outcome of this light-matter interaction, namely a fledgling novel phase with broken time reversal symmetry (TRS) despite $\Delta = 0$, with an extension of our model Hamiltonian.

The paper is organized in the following manner: In Sect. 2, we present the above mentioned simple model of FMWSM bulk with all essential parameters leading to the formation of Weyl node pairs of opposite chirality. We obtain the Berry curvature (BC) and the anomalous Hall conductivity. The latter depends on the angle formed by the spin moments and the axis perpendicular to the plane. In Sect. 3, we report emergence of a novel phase with broken TRS due to the normal incidence of circularly polarized radiation. We discuss some issues related to the Ozawa-Nomura model **[17, 18]** in Sect. 4. The paper ends with very brief concluding remarks in Sect. 5.

## 2. Weyl nodes of opposite chirality

We embark on analysing a simple momentum-space model Hamiltonian and the corresponding electronic band structure (in this section) since physics is about the simplest model that fits the data. Before we do this, it should be mentioned that the two-orbital archetypal model of A. Ozawa and K. Nomura **[17, 18]** represents the ferromagnetic WSM $Co_3Sn_2S_2$ with great details with the effective Kane-Mele type spin-orbit coupling (SOC) between the Co atoms in the kagome plane. The unit cell includes three Co atoms on the Kagome lattice and one Sn atom on the triangular lattice. The Co atoms give rise to the ferromagnetic order. The model is described by a primitive lattice with three basis vectors:

*$a_1$, $a_2$,* and *$a_3$* where $\boldsymbol{a_1} = a\left(\frac{1}{2}, 0, \frac{c}{a}\right)$, $\boldsymbol{a_2} = a\left(-\frac{1}{4}, \frac{\sqrt{3}}{4}, \frac{c}{a}\right)$ and $\boldsymbol{a_3} = a\left(-\frac{1}{4}, -\frac{\sqrt{3}}{4}, \frac{c}{a}\right)$. These authors set $\frac{c}{a} = \frac{\sqrt{3}}{2}$. In the case of bulk $Co_3Sn_2S_2$ the lattice parameters are 5.37 Å and 13.15 Å along (*a, b*) and *c* directions, respectively. The angle α = 59.91647° (58.33°). The angle between directions of *a* and *b* is $\gamma = 118.78° \approx 120°$. We will discuss about the some of the outcomes of this model in section 4. The high symmetry points (HSPs) in the reciprocal space are important for describing the electronic and magnetic properties of solids. Our calculation outcomes of the coordinates of HSPs in the momentum space, setting $\frac{c}{a} = \frac{\sqrt{3}}{2}$, are T(0,0,0.5774), U(0.5556, 0,0.5774), W(−0.5556, 0,0.5774), and L (0.6670, 0, 0.1925). The calculation leans heavily on the article **[17,18]**. We use these results below in graphical representations (with HSPs indicated in the plots) of the single particle excitation spectra of our continuum model. Here one needs to keep in mind that, depending on the conventions used for defining the reciprocal lattice vectors and the choice of the primitive cell, there may be some differences in the exact coordinates of high-symmetry points reported in different studies or databases. However, the relative positions of the high-symmetry points and their symmetry properties should be consistent among different descriptions.

The Weyl point pairs of opposite chiralities are exhibited by FMWSM as mentioned earlier. In momentum space, they act as a paired monopole and anti-monopole of BC. The simple, continuum model of FMWSM in refs. **[20-22]** in the modified form is given by $H_{4B} = [(M(k_x, k_y, k_z) - \xi\zeta)\sigma_z + \xi\hbar v \mathbf{k} \cdot \boldsymbol{\sigma} + \Delta(\eta_1, \eta_2) \cdot \boldsymbol{\sigma}]$. We shall assume below $\hbar = 1$. The model is expected to exhibit the Weyl nodes of opposite chirality feature. Here $M(k_x, k_y, k_z) = M_0 - M_1 a^2(k_x^2 + k_y^2) - M_1 c^2 k_z^2$ with $\mathbf{k} = (k_x, k_y, k_z)$ and $M_0, M_1$ as the relevant material-dependent parameters in units of energy. There is no spin-orbit coupling (SOC) term in the Hamiltonian $H_{4B}$. We have assumed $ck_z = ck_{z_0} = 0.5774$ and $0.1925$ for the graphical representations below. The term $M(k_x, k_y, k_z)$ captures the tunnelling effect.

The Pauli matrices $\boldsymbol{\sigma}$ are acting in the space of bands that make contact at Weyl point pairs. The term $v$ stands for the velocity of the states. We have included the spin magnetic moment (clubbed with the exchange field) $\Delta(\eta_1, \eta_2)$ to take care of the ferromagnetic order. Thus, the model lacks TRS. The directions of the magnetic moments are given by $\Delta(\eta_1, \eta_2, \theta) = \Delta(\eta_1 \sin\theta, \eta_2 \sin\theta, \cos\theta)$, where $\eta_1^2 + \eta_2^2 = 1$. For example, $\eta_1 = 0, 0.8, \pm\frac{\sqrt{3}}{2}, \ldots$ and $\eta_2 = 1, 0.6, \pm\frac{1}{2}, \ldots\ldots$ This general spin structure depends on the angle $\theta$ formed by the spin moments and the axis perpendicular to the plane. The value $\theta = 0$ corresponds to the ferromagnetic order along the z-direction, whereas $\theta = 90°$ corresponds to the in-plane spin order. We shall see that the general structure enables us to analyze the problem of the Weyl node formation in a clearer way. The essential ingredients mentioned above lead to a pair of Weyl nodes as shown in Fig. 1(a) ($ck_{z_0} = 0.5774$) and Fig. 1(b) ($ck_{z_0} = 0.1925$) for $\theta = 0$, $\eta_1 = 0$, and $\eta_2 = 1$. Here we have plotted the four energy eigenvalues of $H_{4B}$, namely $e_j(k_x, k_y, k_z = k_{z_0})$ with $j = (1,2,3,4)$. These are

$$e_j = \pm\left((M \pm vk_{z_0} + \Delta\cos\theta \mp \zeta)^2 + v^2 K^2 + \Delta^2\sin^2\theta \pm 2v\Delta\sin\theta(\eta_1 k_x + \eta_2 k_y)\right)^{1/2} \quad (1)$$

as a function of $ak_x$ for the chemical potential of the fermion number $\mu = 0$. The upper case Greek letter $\mathbf{K} = (k_x, k_y)$. The bands of opposite chirality almost linearly cross each other (with band inversion) at Weyl points along the $k_x$ axis (at $k_x = \pm k_w$) in some parameter ranges. It must be noted that even a slight deviation from linearity implies that the hosted Weyl fermions are not perfectly massless. The Weyl points are located above and below the Fermi level ($E_F = 0$). The points behave as monopoles of BCs with positive and negative chirality. The numerical values of the parameters used in the plots are $\Delta = 1$, $M_0 = 0.23$, $M_1 = 3$, $\mu = 0$, $\zeta = 0.8$, $v = 0.26$, $\theta = 0$, $\eta_1 = 0$, and $\eta_2 = 1$. However, the plots in Fig. 1(c) and (d) are different. While in Fig. 1(c) there is no Weyl point pair, in Fig.1(d) the two points of the pair do not seem to be equidistant from $k_x = 0$. The tentative conclusion is that the in-plane spin order lacks the presence of the Weyl nodes at some points in the Brillouin zone(BZ). We shall seek the confirmation below by examining the Berry curvature. In this paper, except in the case of the light-matter interaction in Sect. 3, and Ozawa-Nomura model in section 4, we choose $\Delta$ to be the unit of energy ($\Delta = 1$). Thus, energy values in Figs. 1, 2, and 3 are dimensionless numbers. Similarly, the wavevector is made dimensionless by multiplying them by lattice parameter $a$.

The intrinsic anomalous Hall conductivity (AHC) may be calculated by integrating the Berry curvature on a $k$-mesh-grid of the Brillouin zone. The expression of AHC is $\sigma_{xy} = -\left(\frac{e^2}{h}\right) \sum_n \int_{BZ} \frac{d^3k}{(2\pi)^3} f(e_n(k_x, k_y, k_{z_0}) - \mu) \Omega_n^z(k_x, k_y, k_{z_0})$ at a given value of $k_z = k_{z_0}$, $\mu$ is the chemical potential of the fermion number, $n$ is the occupied band index, $f(e_n(k_x, k_y, k_{z_0}) - \mu)$ is the Fermi-Dirac distribution and $\Omega_n^z(k_x, k_y, k_{z_0})$ is the $z$-component of the Berry curvature (BC) for the $n$-th band. Here-in-after, we shall drop $k_z = k_{z_0}$ appearing in the argument of various functions above. Upon using the Kubo formula

$$\Omega_n^z(K) = -2\hbar^2 [Im \sum_{m \neq n} (e_n(k_x, k_y) - e_m(k_x, k_y))^{-2}$$

$$\langle n, k_x, k_y | \widehat{v_x} | m, k_x, k_y \rangle \langle m, k_x, k_y | \widehat{v_y} | n, k_x, k_y \rangle]. \quad (2)$$

Here $(k_x, k_y, k_z)$ are the Bloch wave vector components, $e_n(k_x, k_y)$ is the band energy, $|n, k_x, k_y\rangle$ are the Bloch functions of a single band. The operator $\widehat{v_j}$ represents the velocity in the $j$ direction. We recall that, for a system in a periodic potential and its Bloch states as the eigenstates, the identity $\langle m, k_x', k_y' | v_\alpha | n, k_x, k_y \rangle = \left(\frac{1}{\hbar}\right) \left(e_n(k_x', k_y') - e_m(k_x, k_y)\right) \langle m, k_x', k_y' | \frac{\partial}{\partial k_\alpha} | n, k_x, k_y \rangle$ is satisfied. This requires the Heisenberg equation of motion

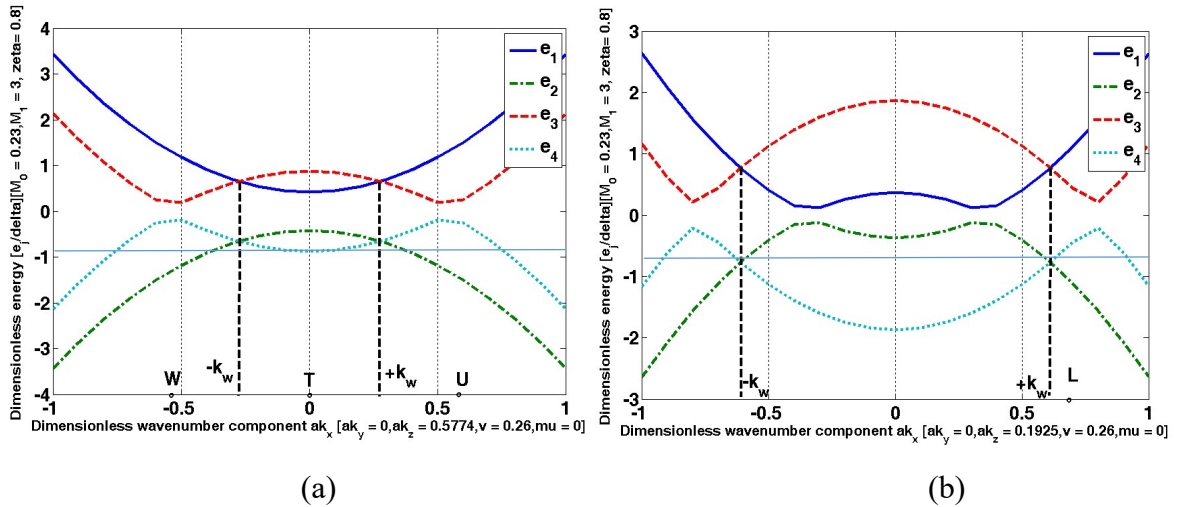

(a)                                                                  (b)

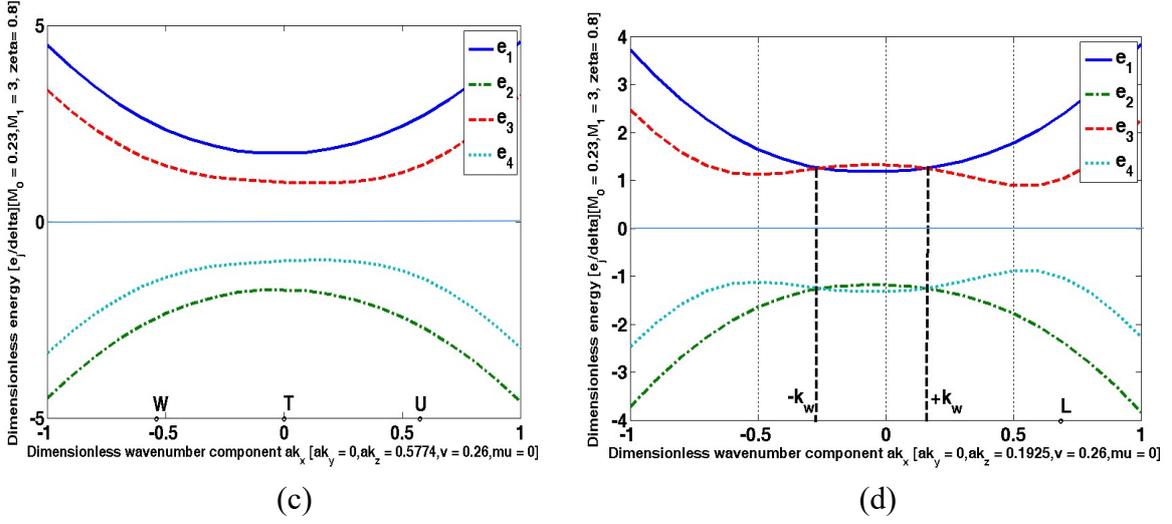

(c)            (d)

**Figure 1.** Plots of the energy eigenvalues in (1) as a function of $k_x$ for $k_y = 0$, $k_{z_0} = 0.5774$(a) and 0.1925 (b). The numerical values of the parameters used are $M_0 = 0.23$, $M_1 = 3$, $\Delta = 1$, $\mu = 0$, $\zeta = 0.8$, $v = 0.26$, $\theta = 0$, $\eta_1 = 0$, and $\eta_2 = 1$. The bands of opposite chirality almost linearly cross each other (with band inversion) at $k_x = \pm k_w$. (c d)The numerical values of the parameters used are $M_0 = 0.23$, $M_1 = 3$, $\Delta = 1$, $\mu = 0$, $\zeta = 0.8$, $v = 0.26$, $\theta = \pi/2$, $\eta_1 = \pm \frac{\sqrt{3}}{2}$, and $\eta_2 = -1/2$. The solid horizontal lines represent the Fermi energy $E_F = 0$.

$i\hbar \frac{d\hat{x}}{dt} = [\hat{x}, \hat{H}]$. Upon using this identity, we obtain Hall conductivity in the zero temperature limit as $\sigma_{xy} = \nu \left(\frac{e^2}{\hbar}\right)$ where $\nu = \sum_n \nu_n$, $\nu_n = \int \int_{BZ} \Omega_{xy}(k_x, k_y) \frac{d^2k}{(2\pi)^2}$. The z-component of BC is $\Omega_{xy}(k_x, k_y) = \left(\frac{\partial A_{n,y}}{\partial k_x} - \frac{\partial A_{n,x}}{\partial k_y}\right) = -2\, Im \left\langle \frac{\partial \psi_{n,K}}{\partial k_x} \middle| \frac{\partial \psi_{n,K}}{\partial k_y} \right\rangle$ where $\psi_{n,K} = |n, k_x, k_y\rangle$. The vector potential $A_n(k_x, k_y)$ is the Berry connection and $\nabla_K \times A_n(k_x, k_y) = \Omega_n(k_x, k_y)$ is the Berry curvature. Since TRS is broken in $Co_3Sn_2S_2$ due to its intrinsic ferromagnetism, it possesses finite values of BC throughout the Brillouin zone (BZ) and sharp peaks at the locations of the Weyl nodes.

In Fig. 2(a) and Fig. 2(b), we have plotted the z-component of BC as a function of $(k_x, k_y)$ for $ck_{z_0} = 0.5774$ and $ck_{z_0} = 0.1925$, respectively. Here the eigenvectors corresponding to eigenvalues in Eq.(1) have been used. We find peaks at the location of nodes. The parameter values in Fig. 2(a) and (b) are $\Delta = 1, \zeta = 0.8, \mu = 0.00, M_0 = 0.23, M_1 = 3, v = 0.10, \theta = 0$, $\eta_1 = 0$, and $\eta_2 = 1$. The Hall conducti-vity is $\sigma_H = 0.8274\left(\frac{e^2}{\hbar}\right)$ in the former case, while in the latter case it is $\sigma_H = 1.0343\left(\frac{e^2}{\hbar}\right)$. The BCs corresponding to the in-plane spin order ($\theta = 90°$) are shown in Fig. 2(c) and (d) for $ck_{z_0} = 0.5774$. In Fig. 2(c), $\eta_1 = 0.8660$, and $\eta_2 = -0.5$, whereas in Fig.2(d) $-\eta_1 = 0.80$, and $\eta_2 = 0.6$. The Hall conductivity is found to be $\sigma_H = 0.0778\left(\frac{e^2}{\hbar}\right)$, and $\sigma_H = 0.0836\left(\frac{e^2}{\hbar}\right)$ for the former and the latter cases, respectively. The non-appearance of sharp peaks in BC in these figures clearly validates the our note above regarding the lack of the presence of the Weyl nodes. The variation in the value of the anomalous Hall-conductivity with the angle formed by the spin moments and the axis perpendicular to the plane is shown in Fig. 2(e) for $\eta_1 = 0.80$, and $\eta_2 = 0.60$.

In order to re-confirm the non-appearance of the sharp-peak aspect, we carry out an investigation with an extension[20−22] of the Hamiltonian $H_{4B}$. The extension, which corresponds to the eight-band Hamiltonian, may be given as $H_{8B} = (M(k_x, k_y, k_z) - \xi\zeta + \xi v c k_z)\tau_0 \otimes \sigma_z + \Delta \cos\theta\, \tau_z \otimes \sigma_0 + \xi v \kappa_x \tau_z \otimes \sigma_x + \xi v \kappa_y \tau_z \otimes \sigma_y$ where the renormalized wavevector components are $\xi v \kappa_x = (\xi v k_x + \eta_1 \Delta \sin\theta)$ and $\xi v \kappa_y = (\xi v k_y + \eta_2 \Delta \sin\theta)$. As before, we shall assume $k_z = k_{z_0} = 0.5774$, and 0.1925 below. The Pauli matrices $\sigma$ and $\tau$ are acting in the space of bands that make contact at Weyl point pairs. It must be mentioned that the Kane-Mele type spin-orbit coupling [17,18,23] is absent in this model. The eigenvalues of this Hamiltonian, for $k_z = k_{z_0}$, are

$$E_i(k_x, k_y, k_{z_0}) = \pm\Delta \cos\theta \pm \left((M + vk_{z_0} - \zeta)^2 + v^2(k_x^2 + k_y^2) + \Delta^2 \sin^2\theta + 2v\Delta\sin\theta(\eta_1 k_x + \eta_2 k_y)\right)^{\frac{1}{2}} \quad (3)$$

for $i = 1,2,3,4$, and

$$E_j(k_x, k_y, k_{z_0}) = \pm\Delta \cos\theta \pm \left[\left((M - vk_{z_0} + \zeta)^2 + v^2(k_x^2 + k_y^2) + \Delta^2 \sin^2\theta\right) - 2v\Delta\sin\theta(\eta_1 k_x + \eta_2 k_y)\right]^{1/2} \quad (4)$$

for $j = 5,6,7,8$. Whereas the single particle energies $E_i$ ($i = 1,2,3,4$) correspond to $\xi = +1$, the energies $E_j$ ($j = 5,6,7,8$) to $\xi = -1$. We have plotted these energy values as a function of $ak_x$ in Figure 3. While there are two pairs of nodes in Figure 3(a)( $ak_{z_0} = 0.5774$) at $k_x = (\pm k_{w_1}, \pm k_{w_2})$ and the band-touching at the T point for $\theta = 0, \eta_1 = 0$, and $\eta_2 = 1$, in Fig. 3(b) ( $ck_{z_0} = 0.1925$) there are four pairs as shown for same values of $(\theta, \eta_1, \eta_2)$. The bands of opposite chirality non-linearly cross each other with the band inversion, at these Weyl points. The absence of linearity implies that the system is unable to host massless Weyl fermions. The parameter values in Fig.3(a) and (b) are $\Delta = 1, \zeta = 0.8, \mu = 0.00, M_0 = 0.23, M_1 = 3$, and $v = 0.10$. The single-particle excitation spectra have been plotted in Fig. 3(c) and Fig. 3(d) for $ck_{z_0} = 0.5774$ and $ck_{z_0} = 0.1925$, respectively, with the same parameter values, but $\theta = \frac{\pi}{2}, \eta_1 = \frac{\sqrt{3}}{2}$, and $\eta_2 = -0.5$. In Fig. 3(c) there is no formed by the Co spin moments and the axis perpendicular to the plane is zero and the factors $\eta_1 = 0$, and $\eta_2 = 1$. (c) $k_z = 0.5774$, $\theta = \pi/2$, $\eta_1 = 0.8660$, and $\eta_2 = -0.5$. (d) $ck_{z_0} = 0.5774, \theta = \pi/2, \eta_1 = 0.80$, and $\eta_2 = 0.6$. The used parameter values are $\Delta = 1, \zeta = 0.8, \mu = 0.019, M_0 = 0.23, M_1 = 3$, and $v = 0.10$. The Hall conductivity is (a) $\sigma_H = 0.8274\left(\frac{e^2}{\hbar}\right)$, (b) $\sigma_H = 1.0343\left(\frac{e^2}{\hbar}\right)$, (c) $\sigma_H = 0.0778\left(\frac{e^2}{\hbar}\right)$, and (d) $\sigma_H = 0.0836\left(\frac{e^2}{\hbar}\right)$. The dashed line in the polar graph represents the variation in the anomalous Hall conductivity value with the angle $\theta$ for $\eta_1 = 0.80$, and $\eta_2 = 0.6$.

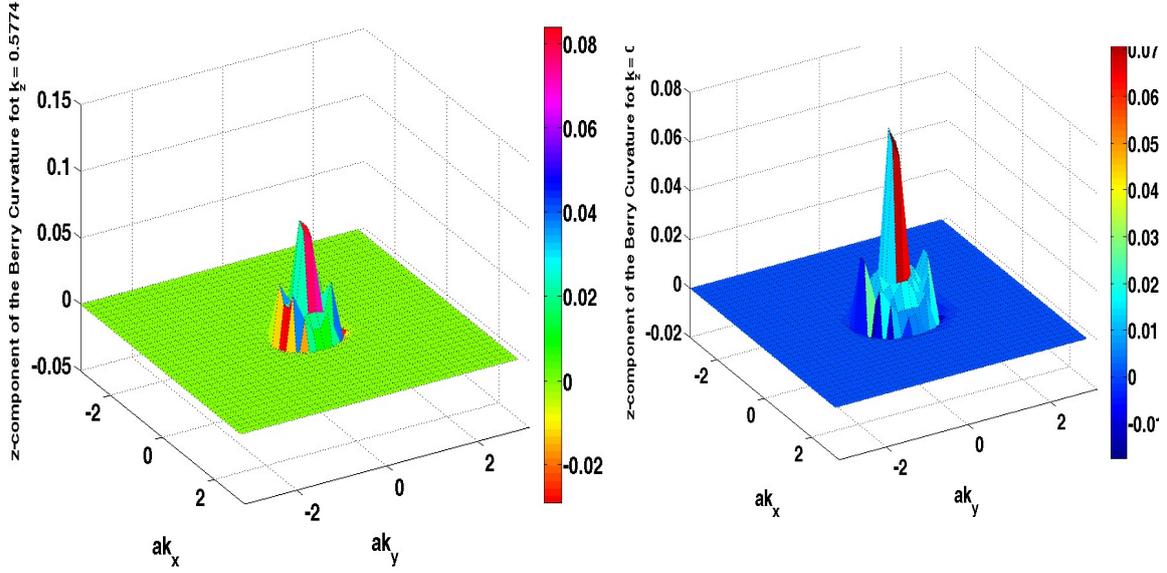

(a) (b)

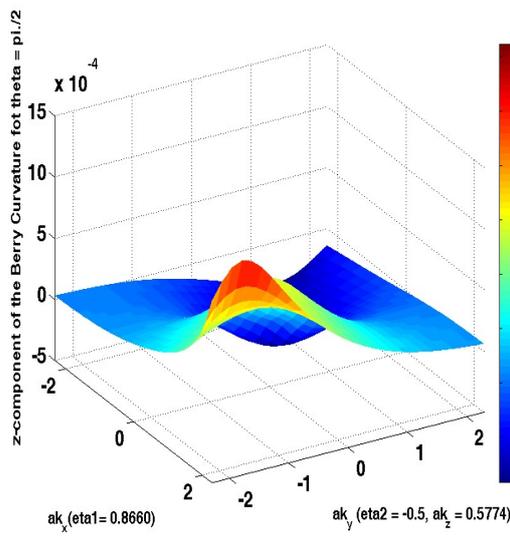
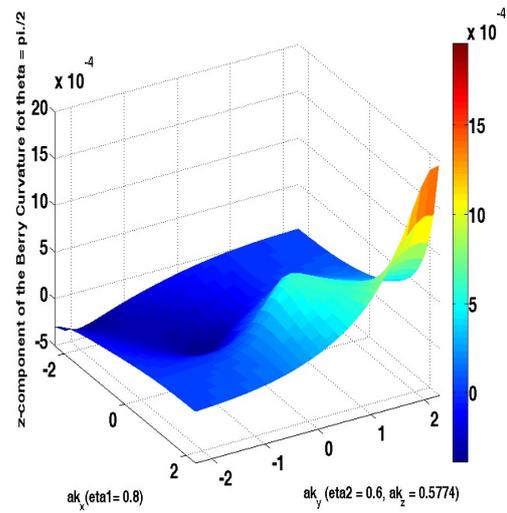

(c) (d)

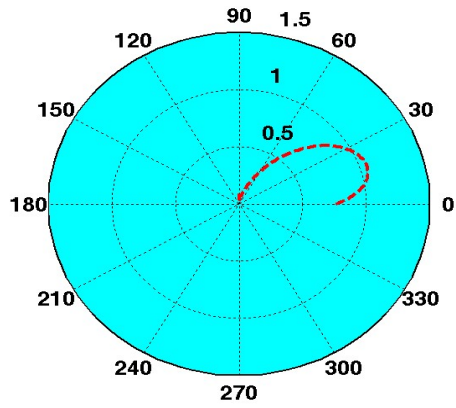

(e)

**Figure 2.** The 3D plots of the z-component of the Berry curvature as a function of ($k_x$, $k_y$). Here the eigenvectors corresponding to eigenvalues in(1) have been used-**(a)** $ck_{z_0}= 0.5774$ and **(b)** $ck_{z_0} = 0.1925$. The angle θ

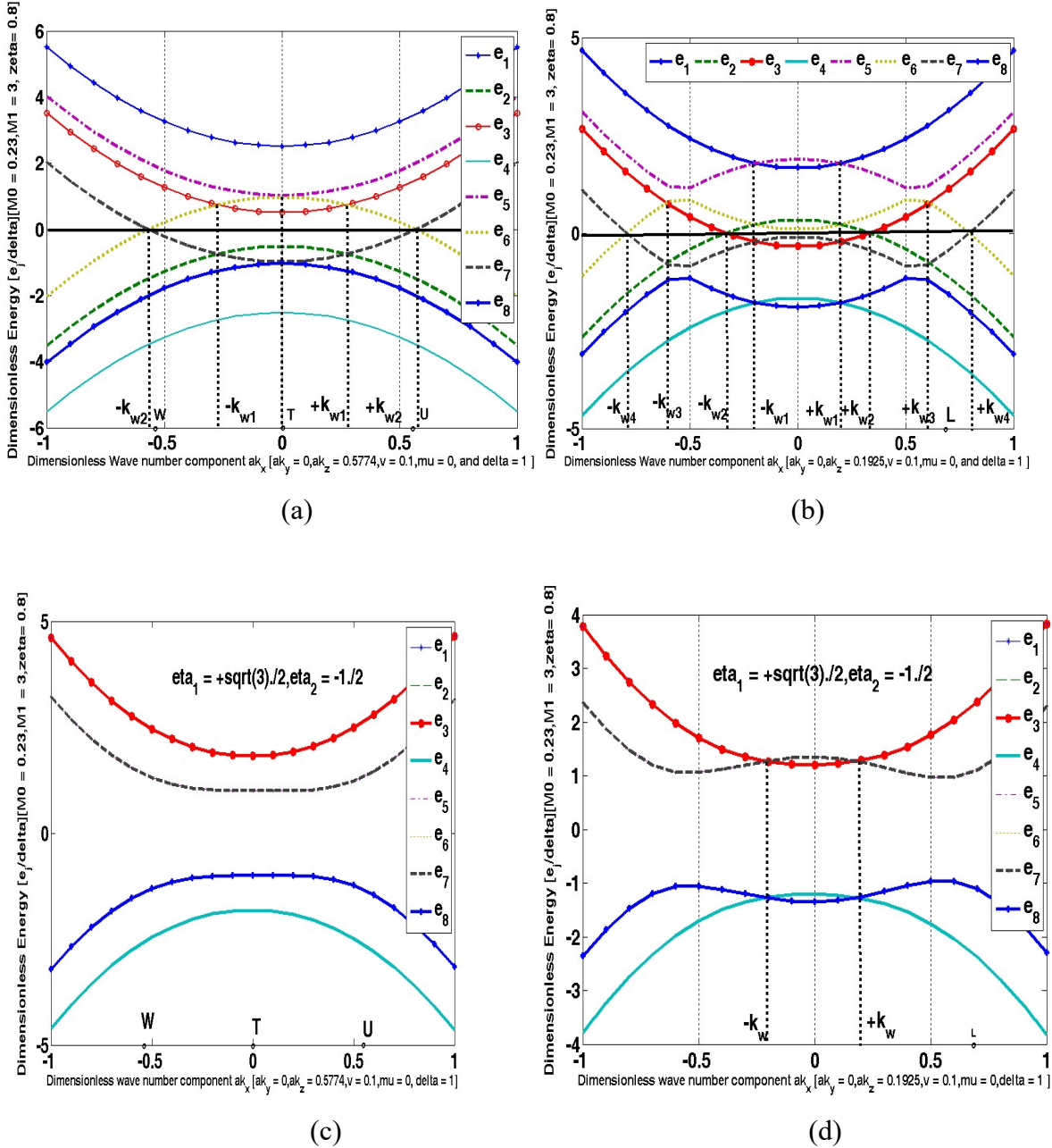

**Figure 3.** Plots of the energy eigenvalues in (3) and (4) as a function of $k_x$ for $k_y = 0$. The numerical values of the parameters used are $M_0 = 0.23, M_1 = 3, \Delta = t_1 = 1, \mu = 0, \zeta = 0.8$, and $v = 0.1$. (a) $ck_{z_0} = 0.5774, \theta = 0, \eta_1 = 0$, and $\eta_2 = 1$; (b) $ck_{z_0} = 0.1925, \theta = 0, \eta_1 = 0$, and $\eta_2 = 1$. The bands of opposite chirality almost linearly cross each other (with band inversion) at $k_x = \pm k_{w_j}$. (c) $ck_{z_0} = 0.5774, \theta = \frac{\pi}{2}, \eta_1 = 0.8660$,

and $\eta_2 = -0.5$. There is no Weyl point pair when (d) $ck_{z_0} = 0.1925, \theta = \frac{\pi}{2}, \eta_1 = 0.8660$, and $\eta_2 = -0.5$. The two Weyl points are equidistant from $k_x = 0$.

Weyl point pair. In Fig. 3(d), however, the two Weyl points are equidistant from $k_x = 0$. There are only four bands in these figures due to the vanishing of the term $\Delta \cos \theta$. Figures 3(c) re-confirms the lack of the presence of the Weyl nodes at some points in the first BZ for the in-plane spin order.

## 3. Interaction of polarized radiation with WSM Film

The radiation-induced response has been found to be very useful in gaining a deep comprehension of the unusual properties of WSM films **[24-26]**. Though some features of Weyl physics in $Co_3Sn_2S_2$ thin films have been investigated **[27-29]**, the surface exposition to circularly polarized radiation (CPR) in the case of FMWSM thin films has been rarely explored so far. We now carry out this investigation for FMWSM with an extension of the model Hamiltonian $H_{8B}$ in Sect.2 by adding the interblock coupling matrix $h_z = \left(\frac{1}{2}\right)[v\,k_z\,(\tau_x + i\tau_y) \otimes \sigma_x + v\,k_z\,(\tau_x - i\tau_y) \otimes \sigma_x]$. In order to obtain surface state Hamiltonian $h_{\text{surface}}$ we make the replacement $ck_z \to -ic\,\partial_z$ and look for states localized within the surface $z = 0$ of the form $exp(-ik_z z)$. We seek such a value of the unknown wave number $ck_z(k_z = -i\chi, \chi > 0)$ for which the exponential $exp(-c\chi(z/c)) \ll 1$ for $z > 0$. For example, if we assume $c\chi \sim 1$ and $\left(\frac{z}{c}\right) \sim 5$, the exponential $exp(-c\chi z/c) \sim exp(-5)$ which is much smaller than unity. Given that the parameter $c \sim 17$ nm for the $Co_3Sn_2S_2$ crystal structure, we obtain $z \sim 85$ nm. Therefore, $k_z = -i\chi, \chi > 0$ ensures decaying term for $z > 0$ in the surface states. The example shows that there is possibility of a fairly localized states at $z = 0$. The surface state Hamiltonian obtained in this manner is of the form
$h_{\text{surface}}(k_x, k_y, \xi, \zeta, \chi) = \vartheta(k_x, k_y, \xi, \zeta, \chi)\tau_0 \otimes \sigma_z + \Delta \cos\theta\, \tau_z \otimes \sigma_0 + \xi v a \kappa_x \tau_z \otimes \sigma_x + \xi v a \kappa_y \tau_z \otimes \sigma_y + h_{z1}$ where $h_{z1} = \left(\frac{1}{2i}\right)[\chi v\,(\tau_x + i\tau_y) \otimes \sigma_x + \chi v\,(\tau_x - i\tau_y) \otimes \sigma_x], \vartheta(k_x, k_y, \xi, \zeta, \chi) = (M(k_x, k_y, \chi) - \xi\zeta - i\,\xi v c\chi)$, and $M(k_x, k_y, \chi) = M_0 - M_1 a^2\,K^2 + M_1 c^2\,\chi^2$ with $\mathbf{K} = (k_x, k_y)$.
The renormalized wave vector components ($\kappa_x, \kappa_y$) are already defined in Sect. 2. As we see below, this form is suitable for investigating the interaction of CPR with FMWSM Film using the Floquet theory.

In the Floquet framework **[30]**, the time-dependent problem could be mapped into effective time-independent formulation analogous to the Bloch theory involving quasi-momentum. We assume the normal incidence of CPR on the surface of FMWSM thin film with the thickness $d = 50$ *nm*. Suppose the optical field of frequency $\omega$ incident on the film is of wavelength $\lambda_{in} \approx 1700\,nm$. Therefore the ratio $d/\lambda_{in} \approx 0.029 \ll 1$. Upon assuming incident optical field as $\mathbf{E}(t) = -\frac{\partial \mathbf{A}(t)}{\partial t} = -\mathbf{E}(cos(\omega t), cos(\omega t + \varphi), 0), \mathbf{E} = \mathbf{A}_1 \omega$, where φ is the phase term and $A_1$ is the amplitude of the corresponding vector potential $A$, we make the Peierls substitution $h_{\text{surface}}(t) = h_{\text{surface}}\left(\mathbf{k} - \frac{e}{\hbar}\mathbf{A}(t)\right)$. The dimensionless quantity $a^2 A_0^2 = \left(\frac{aeE}{\hbar\omega}\right)^2$ corresponds to the light intensity. In particular, when the phase $\varphi = 0$ or $\pi$, the optical field is linearly polarized. When $\varphi = +\pi/2$ ( $\varphi = -\pi/2$), the optical field is left-handed ( right-handed) circularly polarized. Our Hamiltonian $h_{\text{surface}}(t) = h_{\text{surface}}(t+T)$, i.e. it becomes time periodic with the angular frequency $\omega$ and period $T = \frac{2\pi}{\omega}$ upon taking the field into consideration. The Floquet theory can now be applied to our system. The solution of the Schrodinger equation of the system involving Floquet quasi-energy ε appears as $|\varkappa(t)\rangle = \exp(-i\varepsilon t)|\kappa(t)\rangle$ similar to the Bloch formalism. Upon extending this comparability further, one arrives at the fact that the Floquet state satisfies $|\kappa(t)\rangle = |\kappa(t+T)\rangle$ and, therefore, could be expanded in a Fourier series $|\kappa(t)\rangle = \sum_n \exp(-in\omega t)|\kappa_r\rangle$ where $n$ is an integer. Then the wave function, in terms of the quasi-energy ε has the form $|\varkappa(t)\rangle = \sum_n \exp\left(-i\left(\frac{\varepsilon}{\hbar} + n\omega\right)t\right)|\kappa_r\rangle$. This leads to an infinite dimensional eigenvalue equation in the extended Hilbert space[31], namely $\sum_s h_{\text{surface},r,s}|\kappa_n^s\rangle = \left(s\hbar\omega\delta_{r,s} + \frac{1}{T}\int_0^T h_{\text{surface}}(t)e^{i(r-s)\omega t}dt\right)|\kappa_n^s\rangle = \varepsilon_n|\kappa_n^s\rangle$. The matrix element of the Floquet Hamiltonian is now given by $h_{\text{surface},\mu,\nu} = \mu\hbar\omega\delta_{\mu,\nu} + \frac{1}{T}\int_0^T h_{\text{surface}}(t)e^{i(\mu-\nu)\omega t}dt$, where $\mu$ and $\nu$ are integers. These lead to a static effective Hamiltonian, in the off-resonant regime, using the Floquet-Magnus expansion **[32-39]** for high frequency as $h_{\text{surface}}^{Fl}(k) = h_{\text{surface},0,0} + \frac{[h_{\text{surface},0,-1},h_{\text{surface},0,1}]}{\hbar\omega} + O(\omega^{-2})$, where $h_{\text{surface},\mu,\nu} = \frac{1}{T}\int_0^T h_{\text{surface}}(t)e^{i(\mu-\nu)\omega t}dt$ with $\mu \neq \nu$. We now consider the case where $\Delta \ll M_1$. In this case, we obtain

$$h_{\text{surface},0,0} = \vartheta(k_x, k_y, \xi, \zeta, \chi)\,\tau_0\otimes\sigma_z + \xi v k_x \tau_z\otimes\sigma_x + \xi v k_y \tau_z\otimes\sigma_y + h_{z1} - M_1(a^2 A_0^2)\,\tau_0\otimes\sigma_z, \quad (5)$$

$$h_{\text{surface},0,-1} = iM_1(ak_x + e^{-i\varphi}ak_y)aA_0\,\tau_0\otimes\sigma_z - (i/2)\xi(vA_0)\tau_z\otimes\sigma_x - (i/2)\xi(vA_0)e^{-i\varphi}\tau_0\otimes\sigma_y, \quad (6)$$

$$h_{\text{surface},0,1} = -M_1 i(ak_x + e^{i\varphi}ak_y)aA_0\,\tau_0\otimes\sigma_z + (i/2)\xi(vA_0)\tau_z\otimes\sigma_x + (i/2)\xi(vA_0)e^{i\varphi}\tau_0\otimes\sigma_y. \quad (7)$$

Upon using these results, we can write the Floquet Hamiltonian, in units such that $\hbar = 1$, as

$$h_{\text{surface}}^{Fl}(k_x, k_y) =$$

$$\begin{pmatrix} \vartheta_{OP}^+ & \xi A_{OP}^+(ak_-) & 0 & -i\,A_{OP}^+ c\chi \\ \xi A_{OP}^+(ak_+) & -\vartheta_{OP}^+ & -i\,A_{OP}^+ c\chi & 0 \\ 0 & i\,A_{OP}^- c\chi & \vartheta_{OP}^- & -\xi A_{OP}^-(ak_-) \\ i\,A_{OP}^- c\chi & 0 & -\xi A_{OP}^-(ak_+) & -\vartheta_{OP}^- \end{pmatrix}$$

(8)

$$\vartheta_{OP}^\pm(k_x,k_y,\xi,\zeta,\chi) = (M_0 - M_1 a^2 k^2 + M_1 c^2 \chi^2 - \xi\zeta - i\xi vc\chi) + \left(a^2 A_0^2 M_1 \mp \left(\frac{v^2}{a^2}\right)\sin\varphi\right). \quad (9)$$

$$A_{OP}^\pm = \left(\frac{v}{a}\right)\left(1 \mp 2M_1 \sin\varphi\left(\frac{a^2 A_0^2}{\omega}\right)\right), \quad (10)$$

$$k_\mp = k_x \mp i k_y. \quad (11)$$

We now obtain quite an interesting result now. Suppose we drop the renormalized interblock coupling matrix (the counterpart of $h_{z1}$) from Eq.(8). We assume the time reversal operator as $\Theta = -i\tau_0 \otimes \sigma_y K$. We find $\Theta\, h_{surface}^{Fl}(k_x,k_y)\Theta^{-1} = h_{surface}^{Fl}(-k_x,-k_y)$ $+ \left(\frac{4A_0^2 v^2 \sin\varphi}{\omega}\right)\sigma_z \otimes \tau_z$, where $\Theta\, h_{surface}^{Fl}(k_x,k_y)\Theta^{-1} = h_{surface}^{Fl}(-k_x,-k_y)$ when $\varphi = 0$ or $\pi$, that is, when the optical field is linearly polarized. In this case, the time reversal symmetry (TRS) is not broken. However, when $\varphi \neq 0$ or $\pi$, TRS is broken. This result may be interpreted as the emergence of a novel state with broken TRS (despite $\Delta = 0$) due to the incidence of the circularly polarized radiation. The optical field is left-handed circularly polarized for $\varphi = +\pi/2$, whereas it is right-handed for $\varphi = -\pi/2$. One may note that, in the former case, there is a critical intensity of the incident radiation $a^2 A_0^2 \approx \frac{(\hbar\omega)}{2M_1}$ at a given frequency at which the $A_{OP}^+$ will be zero. In the latter case, for the same critical value $A_{OP}^-$ will be zero. This may affect the topological nature of the material. Furthermore, it is worth mentioning that $h_{surface}^{FL}(k_x,k_y,\Delta=0)$ is the Qi-Wu-Zhang (QWZ) model**[40-43]** when the inter-block coupling matrix is $h_{z1}$ dropped in (7). The Hamiltonian $h_{surface}^{FL}(k_x,k_y,\Delta=0)$ corresponds to the quantum spin Hall (QSH) state as shown by these authors. The case, when $h_{z1}$ is retained, needs a detailed investigation.

## 4. Ozawa-Nomura model of Co$_3$Sn$_2$S$_2$

The transition metal-based kagome compound Co$_3$Sn$_2$S$_2$ is a charge transfer metal, where the *p* bands of post-transition metal atoms overlap and hybridize with the transition metal *d* bands near the Fermi level. An archetypal tight binding model was put forward by Ozawa and Nomura **[17, 18]** for the compound Co$_3$Sn$_2$S$_2$. In this model, the intralayer lattice vectors of and the interlayer lattice vectors of kagome layers ( the first-nearest-neighbour vectors) are represented, respectively, by (( *b₁, b₂, b₃* ),( *d₁, d₂, d₃* )), and ( *c₁, c₂, c₃*). This model involves the nearest neighbour, and the next nearest neighbour hopping parameters denoted by $t_{10}$ and $t_2$, respectively, between the orbitals in the Kagome plane. It also involves the hopping integral ($t_{dp}$) between nearest Co and Sn1 sites **[17,18]**, the Kane-Mele type spin-orbit coupling ($t_{soc}$) **[23],** and the nearest neighbour hopping parameter ($t_z$) between the interlayer orbitals. The parameter $t_{10}$ was set as a unit of energy. Thus, energy values in all the figures below are dimensionless numbers. The momentum space representation of the Hamiltonian matrix, in the basis $(d^\dagger_{k.A,\sigma}\ \ d^\dagger_{k,B,\sigma}\ \ d^\dagger_{k,C,\sigma}\ \ p^\dagger_{k.\sigma})^\dagger$ is expressed as

$$\hbar = \begin{pmatrix} -J|m|\sigma_z & A_1 & A_2 & B_1 \\ A_1^* & -J|m|\sigma_z & A_3 & B_2 \\ A_2^* & A_3^* & -J|m|\sigma_z & B_3 \\ B_1^* & B_2^* & B_3^* & \epsilon_p \sigma_0 \end{pmatrix}, \quad (12)$$

$A_1 = -2t_{10}\sigma_0 \cos(\boldsymbol{b_1}.\boldsymbol{k}) - 2t_2\sigma_0 \cos(\boldsymbol{d_1}.\boldsymbol{k}) - 2t_z\sigma_0 \cos(\boldsymbol{c_1}.\boldsymbol{k}) + 2it_{soc}\sigma_z \cos(\boldsymbol{d_1}.\boldsymbol{k})$, (13)
$A_2 = -2t_{10}\sigma_0 \cos(\boldsymbol{b_3}.\boldsymbol{k}) - 2t_2\sigma_0 \cos(\boldsymbol{d_3}.\boldsymbol{k}) - 2t_z\sigma_0 \cos(\boldsymbol{c_3}.\boldsymbol{k}) - 2it_{soc}\sigma_z \cos(\boldsymbol{d_3}.\boldsymbol{k})$, (14)
$A_3 = -2t_{10}\sigma_0 \cos(\boldsymbol{b_2}.\boldsymbol{k}) - 2t_2\sigma_0 \cos(\boldsymbol{d_2}.\boldsymbol{k}) - 2t_z\sigma_0 \cos(\boldsymbol{c_2}.\boldsymbol{k}) + 2it_{soc}\sigma_z \cos(\boldsymbol{d_2}.\boldsymbol{k})$, (15)
$B_1 = -2it_{dp}\sigma_0 \sin\left(\frac{1}{2}\boldsymbol{a_1}.\boldsymbol{k}\right), B_2 = -2it_{dp}\sigma_0 \sin\left(\frac{1}{2}\boldsymbol{a_2}.\boldsymbol{k}\right), B_3 = -2it_{dp}\sigma_0 \sin\left(\frac{1}{2}\boldsymbol{a_3}.\boldsymbol{k}\right)$. (16)

Here $d^\dagger_{k.A,\sigma}$ and $p^\dagger_{k.\sigma}$, respectively, are the creation operators corresponding to d- and p-electrons. The letters (A,B,C) are sub-lattice indices. The $\sigma'_j s$ are the Pauli matrices. Thus, the Hamiltonian $\hbar$ is a $8 \times 8$ matrix. The term ($-J|m|\sigma_z$) describes the ferromagnetic ordering within the mean field approximation, and the energy difference between $p$ orbital and $d$ orbitals is represented by $\epsilon_p$. We need to lean on numerical analysis to obtain the eigenvalues of Eq.(12). We use the 'Matlab' package for this purpose. Here, we choose a $k_j$-path including a high symmetry point. We assume the following dimensionless values of the various parameters: $t_{10} = 1, t_2 = 0.6$, $t_z = -1.0, t_{dp} = 1.80, \epsilon_p = -3.5, t_{soc} = 1.80$ and $J = 2.0$ as in ref.**[17]**. We assign the above-mentioned numerical values to the parameters appearing in the expressions of the matrices ($A_1, A_2, A_3, B_1, B_2, B_3$) in Eqs.(11) to (14). The numerical values of the eigenenergy and the corresponding eigenvectors are obtained using the command [V,D] = eig($\hbar$). The command returns diagonal matrix D of eigenvalues and matrix V whose columns are the corresponding right eigenvectors, so that $\hbar$ *V = V*D. For each k-point, in the chosen $k_j$-path, this process is repeated. The plots of the energy eigenvalues $E_j(k)$ ($j = 1,2,........,8$) as a function of $ak_x$, obtained in this manner, are shown in Figs. 4 and 5 for $ak_y = 0$. In Fig. 4, $ck_z = 0.5774$ and the Kane-Mele type spin-orbit couplings are (a) $t_{soc} = 0$ (b) $t_{soc} = 1.80$. In Fig. 5(a), $ck_z = 0.1925$ and $t_{soc} = 0$, whereas in 5(b) $-ck_z = 0.1925$ and $t_{soc} = 1.80$. The results in Figs. 4 and 5 show that the Weyl point pairs of opposite chiralities together with band inversion (when $t_{soc} = 0$) are conspicuous by their absence, though the avoided crossing features (indicated by verical lines in the plots) in the absence of $t_{soc}$ are in place. When $t_{soc} \neq 0$, the anticrossing feature is present with opening of spectral gaps at some points in the first Brillouin zone(BZ). It must be added that, as shown in the references **[17,18]**, there is emergence of the nodal lines centred at the L point of BZ. This has already been noted in Sect.1.

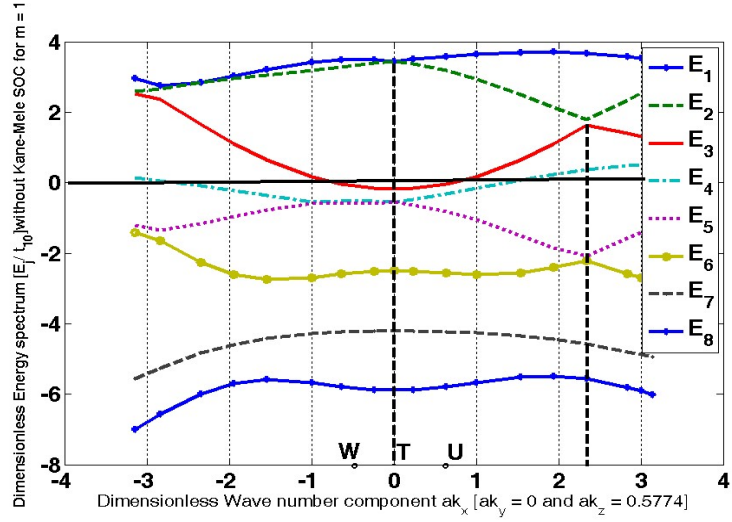

(a)

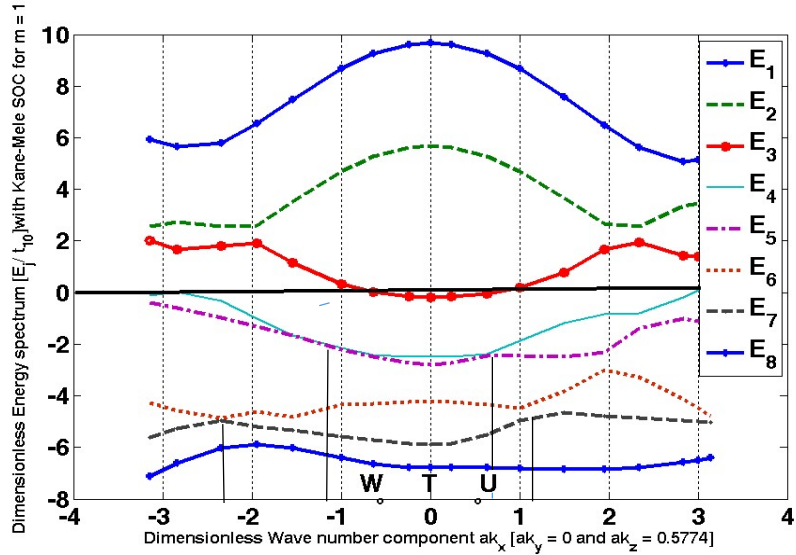

(b)

**Figure 4.** Plots of the dimensionless energy eigenvalues $E_j(k)$ ($j = 1,2,………,8$) corresponding to the matrix in (12) as a function of $ak_x$ for $ak_y = 0,$ and $ck_z = 0.5774$. The numerical values of the various parameters are $t_{10} = 1, t_2 = 0.6$, $t_z = -1.0, t_{dp} = 1.80, \epsilon_p = -3.5,$ and $J = 2.0$ as in ref.**[17]**. In panel (a), $t_{soc} = 0$, while in panel (b) $t_{soc} = 1.80$. The vertical (horizontal) lines indicate avoided crossing momenta ( the Fermi energy $E_F = 0$).

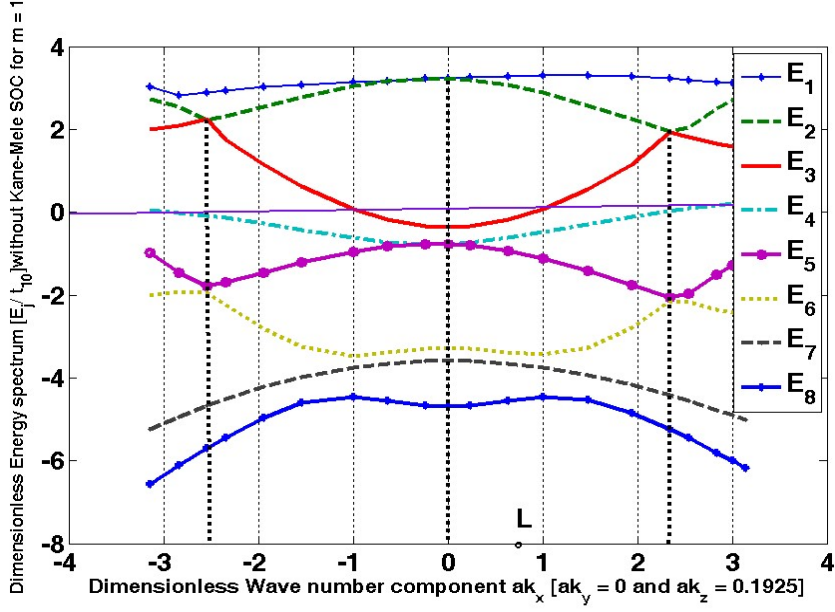

(a)

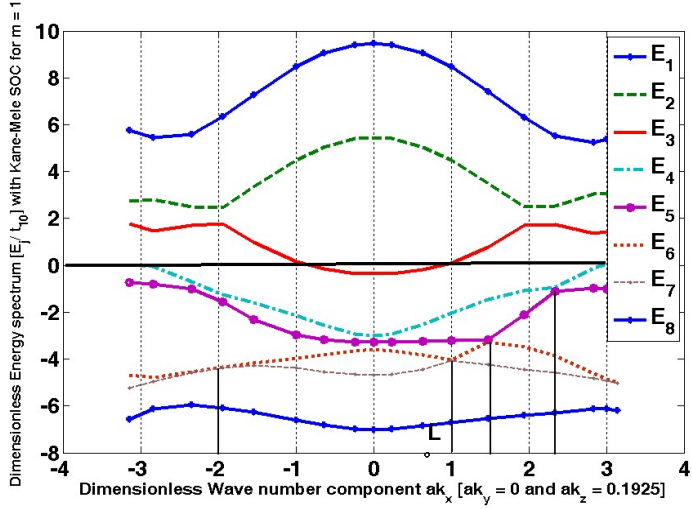

(b)

**Figure 5.** Plots of the energy eigenvalues $E_j(k)$ ($j = 1,2,\ldots,8$) corresponding to the matrix in (12) as a function of $ak_x$ for $ak_y = 0$, and $ck_z = 0.1925$. The numerical values of the various parameters are $t_{10} = 1, t_2 = 0.6$, $t_z = -1.0, t_{dp} = 1.80, \epsilon_p = -3.5$, and $J = 2.0$ as in ref.[17]. In panel (a) $t_{soc} = 0$, while in panel (b) $t_{soc} = 1.80$. The vertical (horizontal) lines indicate avoided crossing momenta (the fermi energy $E_F = 0$).

## 5. Conclusions

Our investigation does not included many details from **[17,18],** as mentioned in Sect. 4. Notwithstanding their absence, the noteworthy feature is that we have been able to show that, for the spin order along the axis perpendicular to the plane of the system, the bands of opposite chirality almost linearly cross each other together with band inversion at Weyl points above and below the Fermi level; for the in-plane spin order, however, the Weyl nodes are absent at some points in the Brillouin zone. Yet another interesting fact is that, strictly speaking, we get the dispersion relations in the non-linear form around the crossing points.

Thus, in this case, massless Weyl fermions are a very remote possibility. In fact, the Weyl fermions with mass seems to be a central feature of the TRS broken WSM **[44,45]**. These are highlights of the present report.

In conclusion, we have also shown the manipulation of topological states of matter by shedding light on them; we found the possibility of emergence of a novel phase with broken TRS ( despite Δ = 0). The novel TRS-broken phase could be established as QSH phase only when one calculates the spin Chern number **[46]**. As in ref.**[46]**, this requires an additional term in system Hamitonian, namely, the Rashba spin-orbit coupling (RSOC) $h_{RSOC} = \left[\left(\frac{1}{2}\right)\left(-\alpha_0 \sin(k_y a)\right)\tau_x \otimes (\sigma_z + \sigma_0) + \left(\frac{1}{2}\right)\alpha_0 \sin(k_x a)\tau_y \otimes (\sigma_z + \sigma_0)\right]$, where $\alpha_0$ stands for the strength of RSOC. It should be mentioned that a new non-centrosymmetric magnetic WSM,i,e. CeAlSi **[47]**, has been identified recently via angle-resolved photo emission spectroscopy. Furthermore, upon substituting Sn with In ($Co_3Sn_{2-x}In_xS_2$),the system under consideration changes from an ferromagnetic WSM to a non-magnetic insulator **[48]**. Zhou et al. **[49]** have observed an enhanced anomalous Hall effect in the compound $Co_3Sn_{2-x}In_xS_2$. Also, the spin Hall conductivity of the Ni-substituted $Co_2Ni_1Sn_2S_2$ exhibits peak-like dependence as shown in **[50].** These problems need extensive investigation. A theoretical study of the compound CeAlSi is our immediate future task. Finally, it is expected that our theoretical investigation will open a new pathway for ultrafast manoeuvring of topological phases in quasi-3D FMWSMs.

---